# EXTINCTION MAP OF BAADE'S WINDOW

K. Z. Stanek[1]

Princeton University Observatory, Princeton, NJ 08544–1001

e-mail: stanek@astro.princeton.edu

## ABSTRACT

Recently Woźniak & Stanek (1996) proposed a new method to investigate interstellar extinction, based on two band photometry, which uses red clump stars as a means to construct the reddening curve. I apply this method to the color-magnitude diagrams obtained by the Optical Gravitational Lensing Experiment (OGLE) to construct an extinction map of $(40')^2$ region of Baade's Window, with resolution of $\sim 30\ arcsec$. Such a map should be useful for studies of this frequently observed region of the Galactic bulge. The map and software useful for its applications are available via anonymous ftp. The total extinction $A_V$ varies from 1.26 $mag$ to 2.79 $mag$ within the $(40')^2$ field of view centered on $(\alpha_{2000}, \delta_{2000}) = $ (18:03:20.9,–30:02:06), i.e. $(l, b) = (1.001, -3.885)$. The ratio $A_V/E(V-I) = 2.49 \pm 0.02$ is determined with this new method.

*Subject headings:* Galaxy: general – stars: Hertzsprung-Russell diagram – stars: statistics

## 1. INTRODUCTION

The Optical Gravitational Lensing Experiment (OGLE, Udalski et al. 1993; 1994) is an extensive photometric search for the gravitational microlensing events towards the Galactic bulge. However, the wealth of data it provides (Szymański & Udalski 1993) may be applied to a host of separate problems. Recently Woźniak & Stanek (1996) proposed a new method for investigating the value of the coefficient of the selective extinction. This method uses red clump stars as a means to construct the reddening curve. Red clump stars are the equivalent of the horizontal-branch stars in metal-rich populations, i.e. they are relatively low-mass stars burning helium in their cores. From observations and also from stellar evolution theory (Castellani, Chieffi, & Straniero 1992) we expect the bulge red clump stars to be relatively bright and have a narrow luminosity distribution with weak dependence on the metallicity. Therefore red clump stars form a suitable population with which to investigate the effects of the interstellar extinction.

---

[1] On leave from N. Copernicus Astronomical Center, Bartycka 18, Warszawa 00–716, Poland

I apply the method of Woźniak & Stanek to the OGLE CMDs to map the interstellar extinction for Baade's Window region of the Galactic bulge. To my knowledge, this is the first attempt to map in detail the extinction in this frequently observed region of the Galactic bulge. Blanco et al. (1984) divided their photographic plate of Baade's Window into four subfields according to the surface density of stars. However, their division was done in rather arbitrary fashion, by visual inspection of the plate. The method I use in this paper, based on Woźniak and Stanek (1996), is entirely quantitative. Terndrup et al. (1995) determined the reddening values for regions defined by Blanco et al. (1984) using medium-resolution spectra of K giants (which, from their color-magnitude diagram, seem overlap largely with what we call "red clump" stars).

In Section 2, I discuss the data used in this paper. In Section 3, I use the method of Woźniak & Stanek to construct the extinction map for nine fields in Baade's Window region of the Galactic bulge. In Section 4, I apply different checks to the results and discuss possible applications of the extinction map.

## 2. THE DATA

I use two somewhat separate data products created by the OGLE collaboration. Szymański & Udalski (1993) constructed a database of photometric measurements in $V$ and $I$ bands for the fields observed by the OGLE project. Udalski et al. (1993) present color-magnitude diagrams of 13 fields in the direction of the Galactic bulge. In this paper I discuss nine fields in Baade's window (BW), which together cover $\sim (40')^2$ and contain $\sim 5 \times 10^5$ stars. All observations were made using the 1-meter Swope telescope at the Las Campanas Observatory (operated by the Carnegie Institution of Washington) and a $2048 \times 2048$ pixel Ford/Loral CCD detector with the pixel size 0.44 *arcsec* covering $15' \times 15'$ field of view. CMDs for the fields analyzed in this paper can be seen in Udalski et al. (1993). Most of each diagram is dominated by bulge stars, with distinct red clump, red giant and turn-off point stars. The part of the diagram dominated by the disk stars has been analyzed by Paczyński et al. (1994). Stanek et al. (1994) has used red clump dominated parts of the CMDs to find a strong signature of the Galactic bar in the OGLE data.

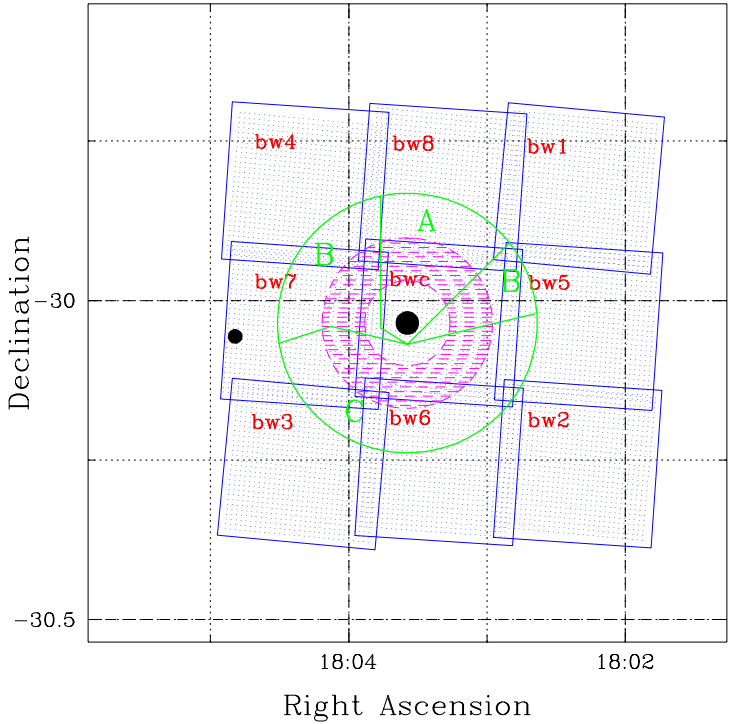

Fig. 1.— Schematic map of Baade's Window 1 deg ×1 deg region of the Galactic bulge. Large overlapping squares represent nine OGLE BW fields. Small dots show the grids on which the extinction was measured. The big circle centered on the large black dot (globular cluster NGC 6522) represents the regions A, B, C observed by Blanco et al. (1984). The shaded annulus centered on NGC 6522 represents the region analyzed by Terndrup et al. (1995). The smaller black dot represents the globular cluster NGC 6528. The coordinates are given in epoch 2000.0.

## 3. THE ANALYSIS

Woźniak & Stanek (1996) describe in detail the method I use in this paper. To summarize, the database (Szymański & Udalski 1993) of $V$ measurements is used to construct star density maps of each of nine Baade's Window fields, BW1-8 and BWC (see Fig.1). Assuming that the extinction affects significantly the number of stars detected in $V$ band in a given subfield, arranging the subfields in order of increasing number of database stars per subfield should reflect a decreasing extinction in a subfield. This indeed has been shown the case by Woźniak & Stanek. The density maps are then used for separating CMDs of subfields with different extinction for a given field. The resulting CMDs are used to obtain the quantitative values of the offset on the CMD between the different subfields, caused by differential extinction. Woźniak & Stanek use the red clump dominated parts of CMDs for determining the offsets – the clump is seen at fainter magnitudes and redder colors in subfields with higher extinction. This method does not depend on knowing the true luminosity function of the clump giants, the only assumption is that it is the same within an $15' \times 15'$ OGLE field. As a result, for each of the nine fields I obtain two independent $30 \times 30$ maps of the relative extinction and reddening within this field, with pixel size corresponding to $64 \times 0.44\ arcsec = 28.16\ arcsec$. The boundary of 64 CCD pixels was removed in each field due to the shifts between $V$ and $I$ frames, resulting in some empty boundary areas.

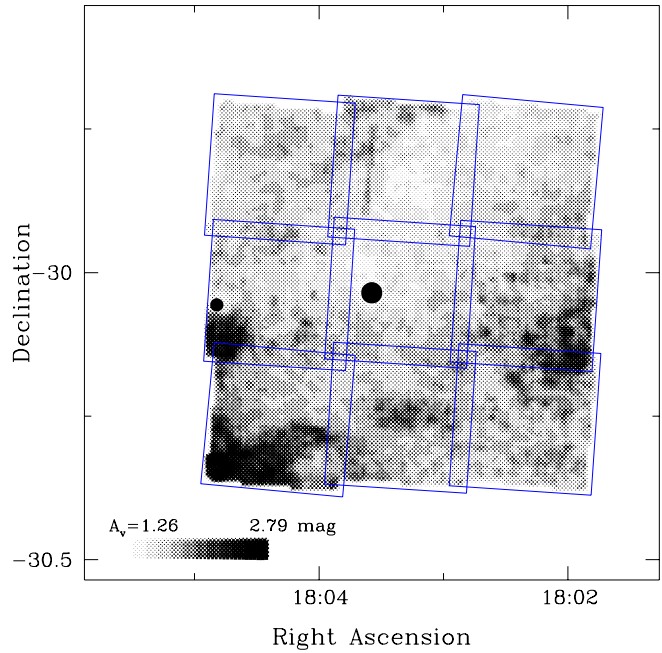

Fig. 2.— Map of the interstellar extinction $A_v$ in Baade's Window. The two black dots correspond to the globular clusters NGC 6522 (center) and NGC 6528 (left). For detailed discussion see text.

Fig. 3.— Image of the same $1\,\text{deg} \times 1\,\text{deg}$ region as in Fig.1 and Fig.2 adapted from the Digitized Sky Survey available on the CD ROMs. Dark regions corresponds to regions of low density of stars, caused by higher extinction. The two globular clusters marked as black dots in Fig.2 can be seen near the center (NGC 6522) and to the left (NGC 6528) of the image.

From the nine $30 \times 30$ maps of the differential extinction and reddening I then create two common maps. First, the individual maps are re-mapped onto one common, uniform grid. As seen in Fig.1, there are overlap regions between the individual fields. These regions are used for matching the values of differential extinction between different fields. They also provide an independent measure of error in our extinction values – one can check how well the extinction and reddening agree in the matching points. Except for the expected zero-point offset the matches are very good, with the rms scatter corresponding to 0.1 $mag$ for the extinction and 0.04 $mag$ for the reddening. This is about twice the formal errors obtained using Woźniak and Stanek method.

So far I have described a method for constructing maps of the differential extinction and reddening within Baade's Window. To construct the final maps, I need zero-points offsets of these variables for some part of the maps. Terndrup et al. (1995) have recently determined $E(V-I)$ for sample of stars in BW and found that the region "A" seen in Fig.1 has an average of $E(V-I) = 0.59 \pm 0.08\ mag$. Using matching points between my map and their region "A" I apply this value to the reddening map. In order to apply the offset in $A_V$, for every point in my maps I obtain $A_V/E(V-I)$ ratio (this is discussed in the next section). The average ratio of 2.49 is then used to multiply the value of $E(V-I) = 0.59\ mag$ of Terndrup et al. (1995) and the resulting zero-point offset of $A_{V,0} = 1.47\ mag$ is then applied to the extinction map. The resulting map of extinction is shown as a greyscale plot in Fig.2. This map can be compared visually with the Palomar Survey plate of the same region seen in Fig.3. The correspondence of the high extinction regions in my map to low density of stars regions on the Palomar plate is striking. In Fig.4 I show the histogram of $A_V$ values for the points on the common grid. Most of BW has the extinction below $A_V = 2.0\ mag$, but there is a tail of high extinction values, mostly in the BW3 field, as sen in Fig.2. The numerical values of both $A_V$ extinction and $E(V-I)$ reddening maps are available through the anonymous ftp service or by request. For the anonymous ftp service, use astro.princeton.edu server, go to stanek/Extinction subdirectory and retrieve the README file with the instructions.

## 4. DISCUSSION

In the previous section I have obtained maps of the interstellar extinction $A_V$ and reddening $E(V - I)$ in roughly $40' \times 40'$ region of Baade's Window. I will now apply several checks to see if these maps are indeed representative of the interstellar extinction in this region.

As a first check, for every point on the common grid (before applying the zero-point offsets) I plot the value of $A_V$ as a function of $E(V - I)$. This is shown in Fig.5. By fitting a straight line to these points, one can obtain the value of $A_V/E(V - I)$, which in our case is $A_V/E(V - I) = 2.49 \pm 0.02$. This is close to the value 2.6 of Dean, Warren, & Cousins (1978) and Walker (1985).

An additional check comes from the overlap regions between the nine BW fields. The fields are matched so that BWC is used to match other four adjacent fields, which in turn are used to match "corner" fields (see Fig.1). So, for each of the corner fields the two overlap regions can be used to see if, for example, matching the BW3 through the BW6 or through the BW7 is producing similar results. I have found no significant difference in the resulting maps regardless of the way the matching procedure went.

I have also used my extinction map to check what is the difference in $A_V$ extinction between regions A, B and C of Blanco et al. (1984). I have found that $A_V$ is $\sim 0.2\ mag$ higher in their region C than in their regions A and B, but there is a $\sim 0.5\ mag$ scatter of $A_V$ in all these regions. One should be cautious when using average extinction – examination of the plates of BW shows that there may be structure in the reddening on scales even smaller than our "pixel" size, which is $\sim 30\ arcsec$.

Another warning for users of the reddening and extinction maps – the regions close to the two globular clusters, NGC 6522 and NGC 6528, in Baade's Window are "contaminated" by the cluster's stars, which may result in extinction values too low for these regions. Indeed, examination of Fig.2 shows somewhat lower extinction around both clusters. Also, there is an artifact of CCD bleeding column visible in the BW8 field as a vertical strip of higher extinction. One should employ caution when using the extinction map in these regions, and perhaps use average extinction from the neighboring regions. The FORTRAN program map.f available on anonymous ftp warns the user when the coordinates fall into these "bad" regions.

I will now turn to some of the possible applications of the extinction and reddening maps produced in this paper. One such application was already explored by Woźniak and Stanek – correcting the CMDs for the differences in the extinction within the field (see their Fig.5). Their dereddened CMD is less diffuse and shows both a much more compact red clump and narrower band of red giants, demonstrating that their method can be used to remove the effects of the differential extinction within the observed field. Another possible application involves removing the effects of interstellar extinction from samples of variable stars. It should be stressed here that my maps show only the total extinction and reddening towards Baade's Window, without any information on how they change with the distance. This problem was addressed to some extent by Arp (1965) and Paczyński et al. (1994). Another possible application of this extinction map would involve measuring the statistical properties of interstellar extinction distribution.

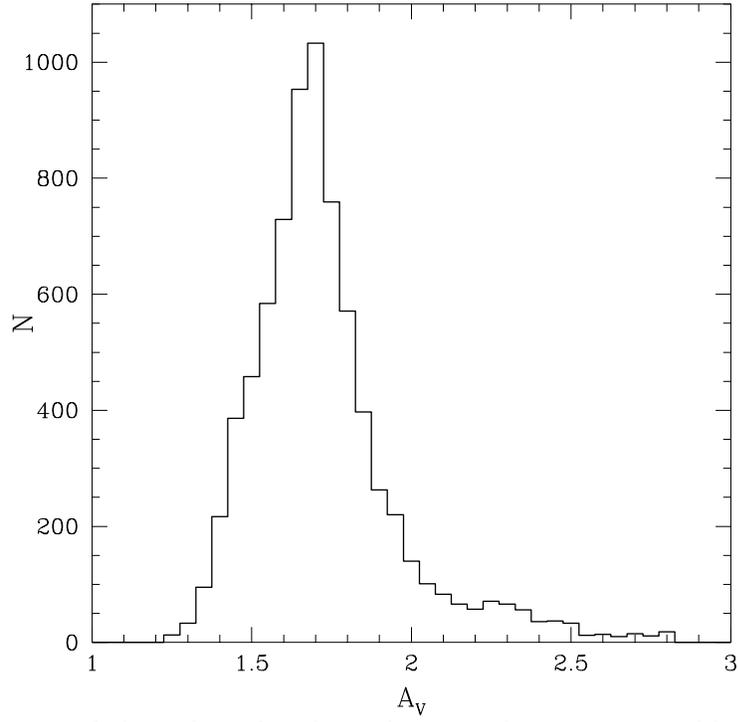

Fig. 4.— Histogram of $A_v$ values for the points on the common grid.

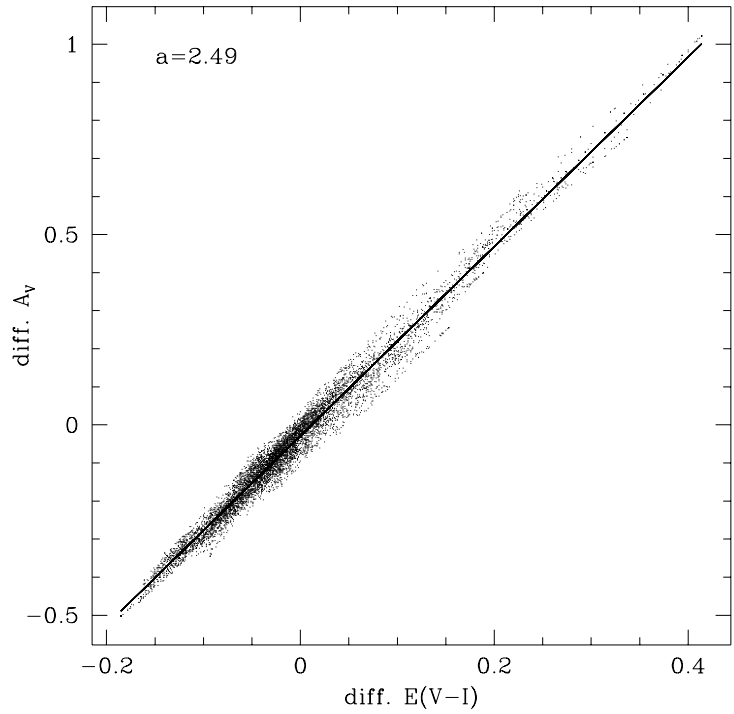

Fig. 5.— Correlation between differential reddening $E(V-I)$ and differential extinction $A_v$.

In this paper I have constructed extinction and reddening maps of Baade's Window region of the galactic bulge, using OGLE CMD data. These maps show highly irregular extinction, with a difference between the lowest and highest extinction regions of more than 1.5 $mag$ in $A_V$. The errors of the maps are dominated by the error of the zero-point offset, taken from Terndrup et al. (1995). When using the map to correct for the differential extinction within BW region, the errors are $\sim 0.1$ $mag$ in $A_V$ and $\sim 0.04$ $mag$ in $E(V-I)$. The numerical values of both $A_V$ extinction and $E(V-I)$ reddening maps are available through the anonymous ftp service or by request.

I would like to thank the OGLE collaboration for making their data available to me and Michał Szymański for his invaluable help with the OGLE software. I thank Przemek Woźniak for some of the software created during our earlier collaboration and used for this project. I also thank Bohdan Paczyński and Slavek Ruciński for indicating the usefulness of this project and comments on the paper. Wes Colley and the referee of the paper Donald Terndrup has also provided me with very useful comments on this paper. This project was supported with the NSF grant AST 9216494 and also with the NAS Grant-in-Aid of Research through Sigma Xi, The Scientific Research Society.